\documentclass{article}
\usepackage[left=1.25in,top=1.25in,right=1.25in,bottom=1.25in,head=1.25in]{geometry}
\usepackage{amsfonts,amsmath,amssymb,amsthm}
\usepackage{verbatim,float,url,enumerate}
\usepackage{graphicx,epsfig,psfrag}
\usepackage{dsfont,bm,color,appendix}
\usepackage{natbib}
\usepackage{pdfpages}
\usepackage[ruled, vlined]{algorithm2e}

\DeclareMathOperator*{\argmax}{arg\,max}

\usepackage{subfig}
\usepackage{hyperref}

\title{On the Use of C-index for Stratified and Cross-Validated Cox Model}
\author{Ruilin Li and Robert Tibshirani \\ Stanford University }

\begin{document}
\maketitle

\begin{abstract}
We develop a baseline-adjusted C-index to evaluate fitted Cox proportional hazard models. This metric is particularly useful in evaluating stratified Cox models, as well as model selection using cross validation. Our metric is able to compare all pairs of comparable individuals in strata or cross validation folds, as opposed to only pairs within the same stratum/folds. We demonstrate through simulations and real data applications that the baseline adjusted C-index is more stable and that it selects better model in high-dimensional $L^1$ penalized Cox regression.

\end{abstract}

\section{Introduction}
Survival analysis \citep{cox1996analysis} often involves predicting time-to-event based on a set of covariates and making inference on them.
Cox model \citep{CoxModel} has been widely used in survival analysis mainly for the following two reasons:
\begin{itemize}
    \item It is very flexible since it allows arbitrary time-varying baseline hazard.
    \item Although it is a semi-parametric model (the baseline function is non-parametric), inference on the covariates parameters can be made without estimating the baseline function \citep{PL}. Therefore it is very easy to fit. This can be achieved by choosing the parameters to maximize the partial likelihood function, which does not depend on the baseline hazard (See section 1.2 for more details).
\end{itemize}
A Cox model fitted using partial likelihood cannot be used to predict survival times since the baseline function is not estimated in that process. In particular we are not able to evaluate the fitted model using metrics such as prediction mean-squared error. In this case, the partial likelihood itself (on the test set) could be a measure of goodness of fit, but it is not very interpretable. Concordance index, or C-index, which can be viewed as a generalization of area under the curve for continuous response variables, evaluates a model using the proportion of pairs of observations where the predicted survival time and the true survival time have the same ordering (concordant). Notably the ordering of survival time could be obtained without the baseline hazard (see section 1.2), so we could compute the C-index for a Cox model fitted with partial likelihood. \\
\\
In a stratified Cox model, where the baseline hazard functions are only same for observations belonging to the same stratum, the ordering of survival time of pairs of individuals from different strata can no longer be reasonably inferred without the baseline functions. To evaluate the C-index for stratified Cox model, one simple option is to only compare pairs from the same stratum, and average over all strata. Similarly, when we use cross validation to fit a Cox model, C-index can be computed by only comparing observations in the same CV fold and average all CV folds. In this paper we show that the C-index obtained this way could be improved by incorporating a baseline hazard estimate from the estimated parameters. With such baseline function estimate we could compute the expected survival time of each individual and compare all pairs in the test set (in the stratified case) or in the training set (in the cross validation case). We call this metric the baseline adjusted C-index.

\subsection{C-Index}
Let $f:\mathcal{X}\mapsto \mathbb{R}$ (or $f(\cdot, U)$ for randomized procedures) be a predictive model, and $(X_i, T_i)$, $(X_j, T_j)$ be a pair of independent realizations of predictor-response tuples coming from the same joint distribution. We define the concordance index, or the C-index of $f$ to be
\begin{equation}
    C_f = P\left[f(X_i) < f(X_j) | T_i < T_j\right] + \frac{1}{2} P(X_i = X_j)
\end{equation}
Given data $\{(X_i, T_i)\}_{i=1}^n$, one can estimate $C_f$ through the following consistent estimator:
\begin{equation}
    \hat{C}_f = \frac{\sum_{i,j=1}^n 1[f(X_i) < f(X_j)] 1(T_i < T_j)}{\sum_{i,j=1}^n1(T_i < T_j)} + \frac{\left[\sum_{i,j=1}^n 1(X_i = X_j)\right] - n}{2(n^2 - n)}. 
\end{equation}
To simplify the notation, in the following we focus on the situation where $X$ has a continuous distribution. In particular $P(X_i = X_j) = 0$, and $C_f, \hat{C}_f$ can be written as:
\begin{equation}
    C_f = P\left[f(X_i) < f(X_j) | T_i < T_j\right]
\end{equation}

\begin{equation}
     \hat{C}_f = \frac{\sum_{i,j=1}^n 1[f(X_i) < f(X_j)] 1(T_i < T_j)}{\sum_{i,j=1}^n1(T_i < T_j)} \label{Cest}
\end{equation}
In this paper we focus on applications to survival analysis. It is often the case that the survival time is right-censored, so we augment the predictor with a variable $O \in \{0,1\}$. If $O = 0$, then we only know that the survival time is greater than $T$, otherwise the survival time is exactly $T$. In addition, when censoring happens it is possible for a pair of observations $(T_i, T_j)$ to be incomparable. This happens when:
\begin{itemize}
    \item Both $i$ and $j$ are censored. That is $O_i = O_j = 0$.
    \item $i$ is censored, $j$ is uncensored, but $T_j > T_i$. Or vice versa.
\end{itemize}
In both cases above we do not count that pair in our comparison, and we abuse the notation to set $1(T_i < T_j) = 1(T_j < T_i) = 0$.

\subsection{Cox Proportional Hazard Model}
Given a numerical predictor $X \in \mathbb{R}^d$, Cox model assumes there exists $\beta \in \mathbb{R}^d$ and a baseline function $h_0:\mathbb{R}^+ \mapsto \mathbb{R}^+$ such that the corresponding hazard function for survival time is:
\begin{equation}
    h(t|X) = h_0(t) \exp(\beta^T X) \label{hazard}
\end{equation}
In other words, the cumulative distribution function of the survival time $T$ given $X$ is:
\begin{equation}
    P(T \le t|X) = 1 - \exp\left(- e^{\beta^T X}  \int_0^t h_{0}(s) ds\right)
\end{equation}
One advantage of Cox model is that, while having the flexibility to have any baseline function, we don't need to perform non-parametric estimation on it to make inference on $\beta$. This can be achieved by maximizing the partial likelihood function with respect to $\beta$:
\begin{equation}
    \hat{\beta} = \argmax_{\beta} \sum_{i:O_i = 1} \beta^T X_i -\log \left(\sum_{j:T_j \ge T_i} \exp(\beta^T X_j)\right)
\end{equation}
Since the baseline function $h_0$ is not estimated in this procedure, we can't make predictions on the survival time. However it is till possible to evaluate the model
$h(t|X) = h_0(t) \exp(\hat{\beta}^T X)$ through C-index with the following observation: if $\beta^Tx_i > \beta^T x_j$, then
\begin{equation}
    P(T_i > t | X_i = x_i) < P(T_j > t| X_j = x_j) \text{ for all } t > 0 \label{stogreat}
\end{equation}
That is, if $\beta^T x_i > \beta^T x_j$ then $Y_j$ is stochastically greater than $Y_i$. In particular, the mean, median, and all quantiles of $Y_j$ are greater that that of $Y_i$. Therefore it is reasonable to define the C-index in this case by replacing $f(X)$ in \eqref{Cest} with $-\hat{\beta}^T X$:

\begin{equation}
    \hat{C}(\hat{\beta}) = \frac{\sum_{i,j=1}^n 1[\hat{\beta}^T X_i > \hat{\beta}^T X_j] 1(T_i < T_j)}{\sum_{i,j=1}^n1(T_i < T_j)} \label{basicC}
\end{equation}
In a stratified Cox model, we assumes data comes from $K$ different populations with heterogeneous baseline hazard but the same $\beta$. In particular, the hazard function for the $k$-th stratum is given by:
\begin{equation}
    h_k(t|X) = h_{0,k}(t)\exp(\beta^TX), \quad k = 1,2,\cdots, K \label{stratCox}
\end{equation}
We further augment the predictor $X_i$ with a stratum index $S_i \in \{1,\cdots, K\}$ indicating which stratum the $i$th individual comes from. In this case, inference on $\beta$ could still be performed using the partial likelihood function, which in the stratified model becomes:
\begin{equation}
    \hat{\beta} = \argmax_{\beta} \sum_{k=1}^K \sum_{i:O_i = 1, S_i=k} \beta^T X_i -\log \left(\sum_{j:T_j \ge T_i, S_j=k} \exp(\beta^T X_j)\right) \label{beta}
\end{equation}
Again the baseline functions $h_{0,k}$ are not estimated in this procedure. To evaluate $\hat{\beta}$, we would like to use the C-index estimation similar to \eqref{basicC}. However, in this case the observation \eqref{stogreat} only holds true if $S_i = S_j$. As a result we are only justified to compare pairs from the same stratum, which leads to the C-index estimate:
\begin{equation}
    \hat{C}(\hat{\beta}) = \frac{1}{K}  \sum_{k=1}^K   \frac{\sum_{i,j=1}^n 1[\hat{\beta}^T X_i > \hat{\beta}^T X_j] 1(T_i < T_j)1(S_i = S_j = k)}{\sum_{i,j=1}^n1(T_i < T_j)1(S_i = S_j = k)} \label{C0}
\end{equation}
Individuals from different strata are not compared in this C-index estimate.

\section{Method}
\subsection{Stratified Cox Model}
In this section we define the baseline adjusted C-index for stratified Cox proportional hazard model, denoted as $\hat{C}_{ba}(\hat{\beta})$. Later we will demonstrate the usage of $\hat{C}_{ba}$ in cross-validation.\newline
\\
One drawback of the C-index estimate \eqref{C0} is that pairs across different strata are not compared. To remedy this problem, given a parameter estimate $\hat{\beta}$, we first produce a non-parametric cumulative baseline function estimate using $\hat{\beta}$:
\begin{equation}
    \hat{H}_{0, k}(t) = \sum_{i: T_i \le t, S_i =k}  \frac{O_i}{\sum_{j:S_j=k, T_j \ge T_i} \exp(\hat{\beta}^T X_j)} \label{base}
\end{equation}
where the true cumulative hazard function is defined to be
\begin{equation}
    H_{0,k}(t) = \int_{0}^t h_{0,k}(s) ds
\end{equation}
One can show that $\hat{H}_{0,k}(t)$ is a consistent estimator for $H_{0,k}(t)$ for all $k \in [K]$, $t \ge 0$, provided that $\hat{\beta}$ is consistent. With the baseline estimate \eqref{base}, we could obtain a survival time prediction using the expected survival time:
\begin{align}
    \begin{split}
        \mathbb{E}(T|X=x, S = k) &= \int_{t=0}^\infty  \exp\left(-\int_0^t h_{0,k}(s) ds e^{\beta^T x}\right)dt
        \\ &= \int_{t=0}^\infty \exp\left(-H_{0,k}(t) e^{\beta^T x}\right) dt
    \end{split}
\end{align}
Since the cumulative baseline estimate \eqref{base} is a step function that only changes at event time, we could approximate the above integral numerically using trapezoidal rule to obtain a predicted survival time $\hat{T}$ given $X=x, S=k$. To do this we first sort the events time in the $k$th stratum $0 = T^k_{(0)} \le T^k_{(1)} \le \cdots \le T^k_{(n_k)}$, where to simplify notation we add a pseudo-observation $T^k_{(0)}$ and $n_k$ is the number of individuals in stratum $k$. The predicted survival time given $X=x, S=k$ can be written as
\begin{equation}
    \hat{T} = \sum_{i=1}^{n_k} \frac{T^k_{i} - T^k_{i-1}}{2} \left\{\exp\left[- e^{\hat{\beta}^Tx}\hat{H}_{0, k}(T^k_{i})\right] + \exp\left[- e^{\hat{\beta}^Tx}\hat{H}_{0, k}(T^k_{i-1})\right] \right\} \label{pred}
\end{equation}
where $\hat{H}_{0,k}(0) = 0$. Let $\mathcal{D}^{test} = \{(X_i^{test}, O_i^{test}, S_i^{test}, T_i^{test})\}_{i=1}^m$ be the test set where we would like to evaluate the C-index of $\hat{\beta}$. We define the baseline adjusted C-index on the test set to be
\begin{equation}
    \hat{C}_{ba}(\hat{\beta}) =   \frac{\sum_{i,j=1}^m 1[\hat{T}_i^{test} < \hat{T}_j^{test}] 1(T_i^{test} < T_j^{test})}{\sum_{i,j=1}^n1(T_i^{test} < T_j^{test})} \label{stratC}
\end{equation}
where $\hat{T}_i^{test}$ is computed using \eqref{pred}. A pseudo-code to compute $\hat{C}_{ba}$ is given in the next page.

\begin{algorithm}[H]
\SetAlgoLined
 \SetKwInOut{Input}{Input}
 \Input{$\mathcal{D}^{train} = \{(X_i, O_i, S_i, T_i)\}_{i=1}^n$
 \\$\mathcal{D}^{test} = \{(X_i^{test}, O_i^{test}, S_i^{test}, T_i^{test})\}_{i=1}^m$ }
\KwResult{Baseline adjusted C-index}

 Estimate $\hat{\beta}$ on $\mathcal{D}^{train}$ using partial likelihood function as in \eqref{beta}\;
 Estimate the baseline functions $\hat{H}_{0,1},\cdots,\hat{H}_{0,K}$ on $\mathcal{D}^{train}$ as in \eqref{base}\;
 Predict survival time of individuals in test set $\{\hat{T}_i^{test}\}_{i=1}^m$ using \eqref{pred}\;
 Computer $\hat{C}_{ba}(\hat{\beta})$ using \eqref{stratC} \;
\Return{$\hat{C}_{ba}(\hat{\beta})$}
\caption{Baseline adjusted C-index for Stratified Cox model}
\end{algorithm}

\subsection{Cross Validation in Cox Model}
Similar ideas from the last section can be applied for model selection. Here we will focus on $L^1$-regularized \cite{LASSO} Cox regression . The described procedure can be generalized to other families of Cox models directly.\\
\\
$L^1$-regularized Cox regression selects a model and estimate model parameters simultaneously by maximizing the weighted sum of the Cox partial likelihood and the $L^1$ norm of the parameters:
\begin{equation}
    \hat{\beta}(\lambda) = \argmax_{\beta} - \lambda \|\beta\|_1 + \sum_{i:O_i = 1} \beta^T X_i -\log \left(\sum_{j:T_j \ge T_i} \exp(\beta^T X_j)\right)  \label{reg}
\end{equation}
In practice, the hyper-parameter $\lambda$ is often selecting through cross validation, where the training set is split randomly into $K$-folds (common choices of $K$ are $K = 2, 5, 10$). After the splitting, we train the model on the training set consisting of $K-1$ of the $K$ folds. One option to evaluate a certain $\lambda$ is to compute the regular C-index on the validation set using the model trained on the training set, and then average over all validation sets. Here we propose the baseline adjusted C-index for cross validation where for each training set we first estimate $\hat{\beta}$ and use it to obtain a baseline function estimate defined in \eqref{base} (here $K = 1$). With $\hat{\beta}$ and $\hat{H}_{0}$ we could predict the survival time $\hat{T}_i$ as \eqref{pred} for each observations in the validation set. Repeat this step for each training-validation split and then we define the baseline adjusted C-index for $\lambda$ to be the proportion of concordant pairs among all observations, using the predicted survival time. A pseudo code for this procedure is provided here.

\begin{algorithm}[H]
\SetAlgoLined
 \SetKwInOut{Input}{Input}
 \Input{$\mathcal{D}^{train} = \{(X_i, O_i, S_i, T_i)\}_{i=1}^n$;
 \\ number of cross validation folds $= K$; \\ a sequence of candidate regularization parameter \texttt{lambda\_seq} }
\KwResult{Best regularization parameter in \texttt{lambda\_seq} in terms of baseline adjusted C-index}

Randomly split $\mathcal{D}^{train}$ into $K$ disjoint subsets of equal number of observations $\mathcal{D}^{train} = (\mathcal{D}^{1}, \cdots, \mathcal{D}^{K})$\;
Initialize a vector $\hat{C}_{ba}$ to store the baseline adjusted C-index for each regularization parameter\;

\For{$\lambda$ in \texttt{lambda\_seq}}{
    \For{$k$ in $1:K$}{
        Set $\mathcal{D}^{-k} = \{\mathcal{D}^i: i \neq k\}$\;
        Estimate $\hat{\beta}$ on $\mathcal{D}^{-k}$ using penalized partial likelihood with parameter $\lambda$ as in \eqref{reg}\;
        Estimate the baseline functions $\hat{H}_{0,1},\cdots,\hat{H}_{0,K}$ on $\mathcal{D}^{train}$ as in \eqref{base}\;
        Predict survival time of individuals in the validation set $\mathcal{D}^k$ using \eqref{pred}\;
        Set $\hat{T}_i$ to be the predicted survival time for all individuals $i$ in $\mathcal{D}^k$ \;
    }
    Use computed $\{\hat{T}_i\}_{i=1}^n$ to obtain a C-index using formula \eqref{Cest} with $f(X_i) = \hat{T}_i$. Store the result to $\hat{C}_{ba}(\lambda)$
}
\Return{$\argmax \hat{C}_{ba}$}
\caption{Baseline adjusted C-index in cross validated $L^1$ regularized Cox regression model}
\end{algorithm}




\section{Simulation}
\subsection{Stratified Cox Model}
To demonstrate the stability (low variance) of the baseline adjusted C-index. We first simulated stratified survival data where the predictors $X_1,\cdots, X_n$ are i.i.d multivariate normal random variables, $S_i \in \{1,2,\cdots, 10\}$ (10 strata).  Conditional on $X_i, S_i$, the survival time $T_i$ follows the stratified Cox model exactly with constant baseline functions. In this example, the survival time is not censored. We split the data into training and test set. On the training set, we estimate $\hat{\beta}$ using partial likelihood. On the test set, we compute both the C-index that only compares pairs in the same strata \eqref{C0}, the C-index that compares all pairs based on $\hat{\beta}^TX$, and the baseline adjusted C-index \eqref{stratC}. We simulate this process 200 times. \ref{Figstrat} is a box plot of the computed C-indices:
\begin{figure}%
    \centering
    \subfloat[Regular Case]{{\includegraphics[width=0.45\linewidth]{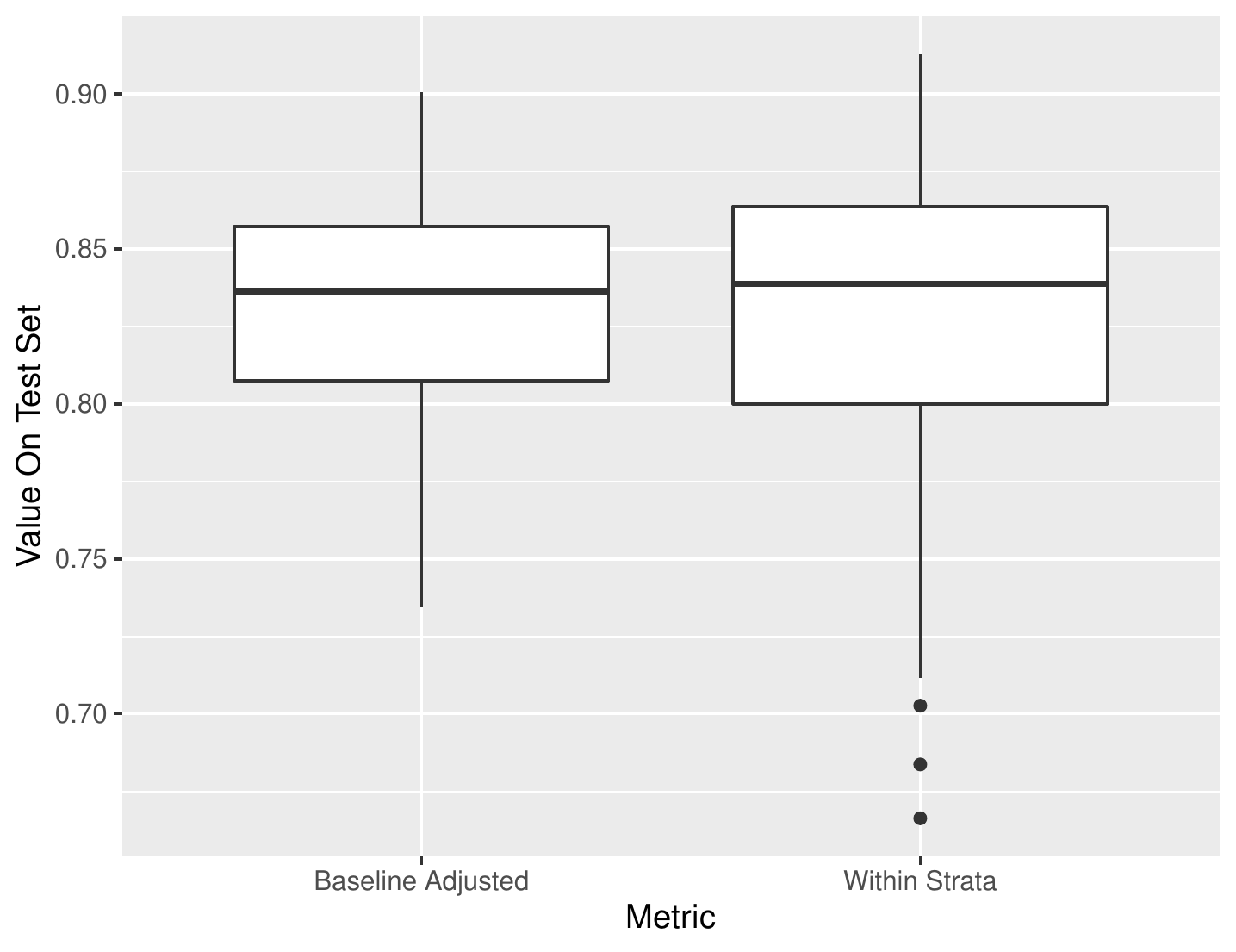} }}%
    \qquad
    \subfloat[Low Signal Case]{{\includegraphics[width=0.45\linewidth]{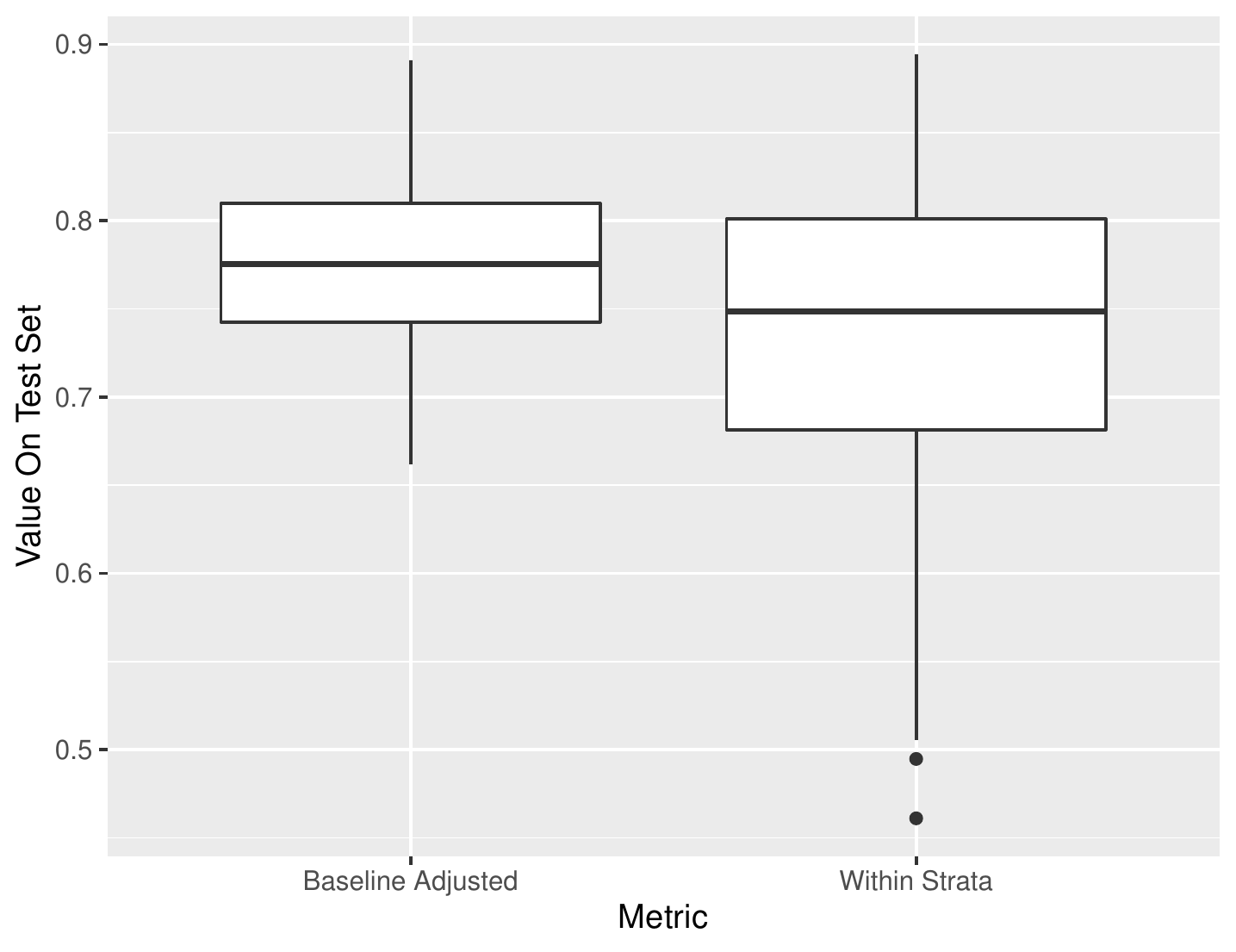} }}%
    \caption[]{\em Comparison between two metrics in stratified Cox model. Left plot: we fit Cox models using partial likelihood on the training set and evaluate both the baseline adjusted C-index and the default C-index that only compares individuals in the same strata. The box plots are based on 200 simulations. The right plot: we lower the predictive power of the covariates (low signal) by setting $\beta$ to be very sparse and repeat the experiment for the left plot. }%
    \label{Figstrat}
\end{figure}
One could see that the baseline adjusted C-index is indeed more stable and has less outliers, and the advantage of baseline adjusted C-index is even higher when the signal from the covariates is low.

\subsection{Cross Validation}
In this section we empirically analyze the model selection performance of baseline adjusted C-index in the context of $L^1$ regularized Cox model. Again we assume a constant baseline function, and we use cross validation to find the $\lambda$ in \eqref{reg} that maximizes a validation metrics. The metrics we use include baseline adjusted C-index, within fold C-index, and partial likelihood. The data is simulated under the following settings:
\begin{itemize}
    \item $X_i$ have independent coordinates, model correctly specified.
    \item The coordinates of $X_i$ follows a first order autoregressive process, or an AR(1) process, model correctly specified.
    \item $X_i$ have independent coordinates, the linear part of \eqref{hazard} contains an interaction term that is not fitted in our model (mis-specified model).
    \item The coordinates of $X_i$ follows an AR(1) process, he linear part of \eqref{hazard} contains an interaction term that is not fitted in our model (mis-specified model).
\end{itemize}
Here are the comparison of the metrics in terms of the mean squared error (MSE) of the estimated $\hat{\beta}$ using $\lambda$ that is selected by that metric. 
\begin{figure}%
    \centering
    \subfloat[Regular Case]{{\includegraphics[width=0.45\linewidth]{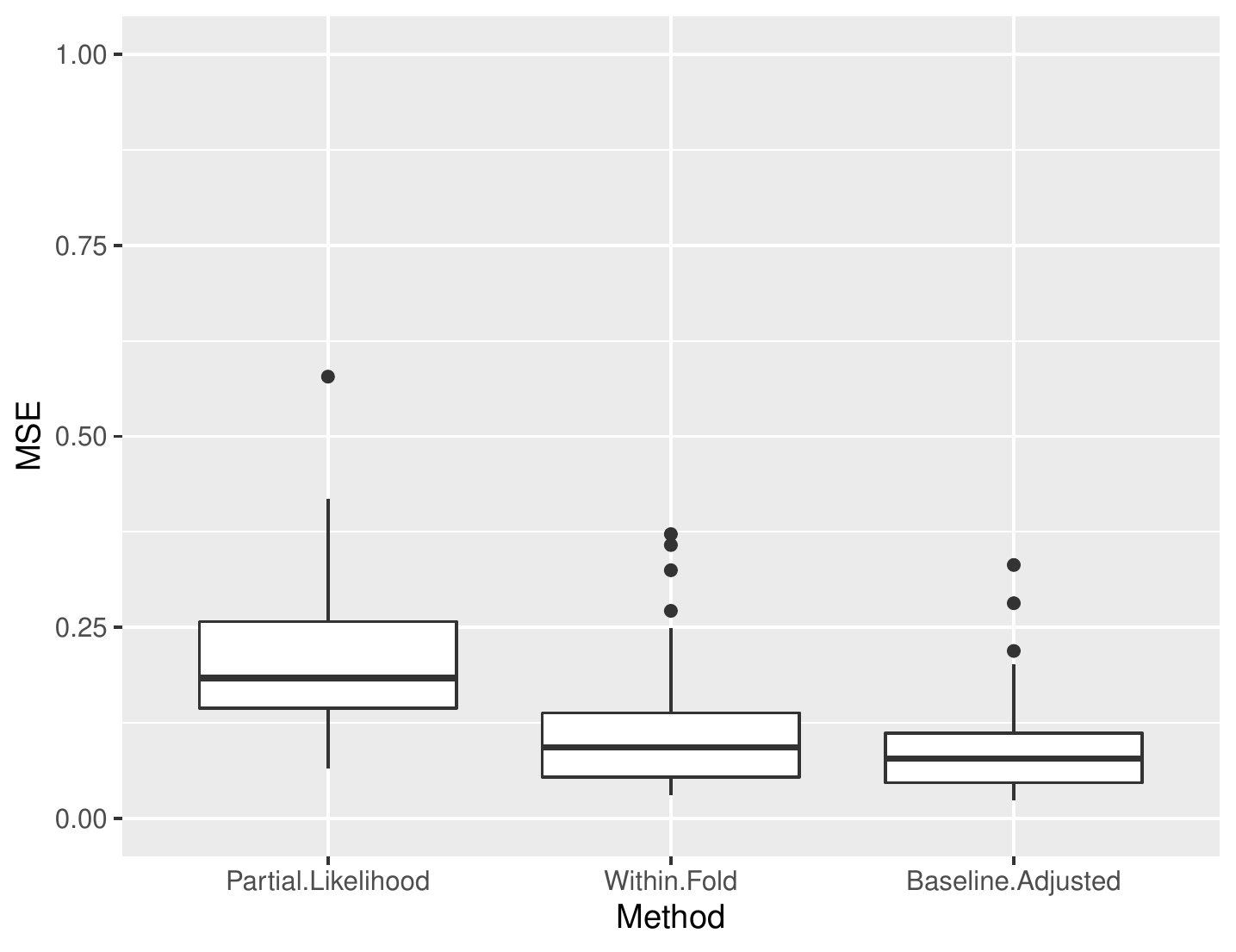} }}%
    \qquad
    \subfloat[AR(1) Predictors]{{\includegraphics[width=0.45\linewidth]{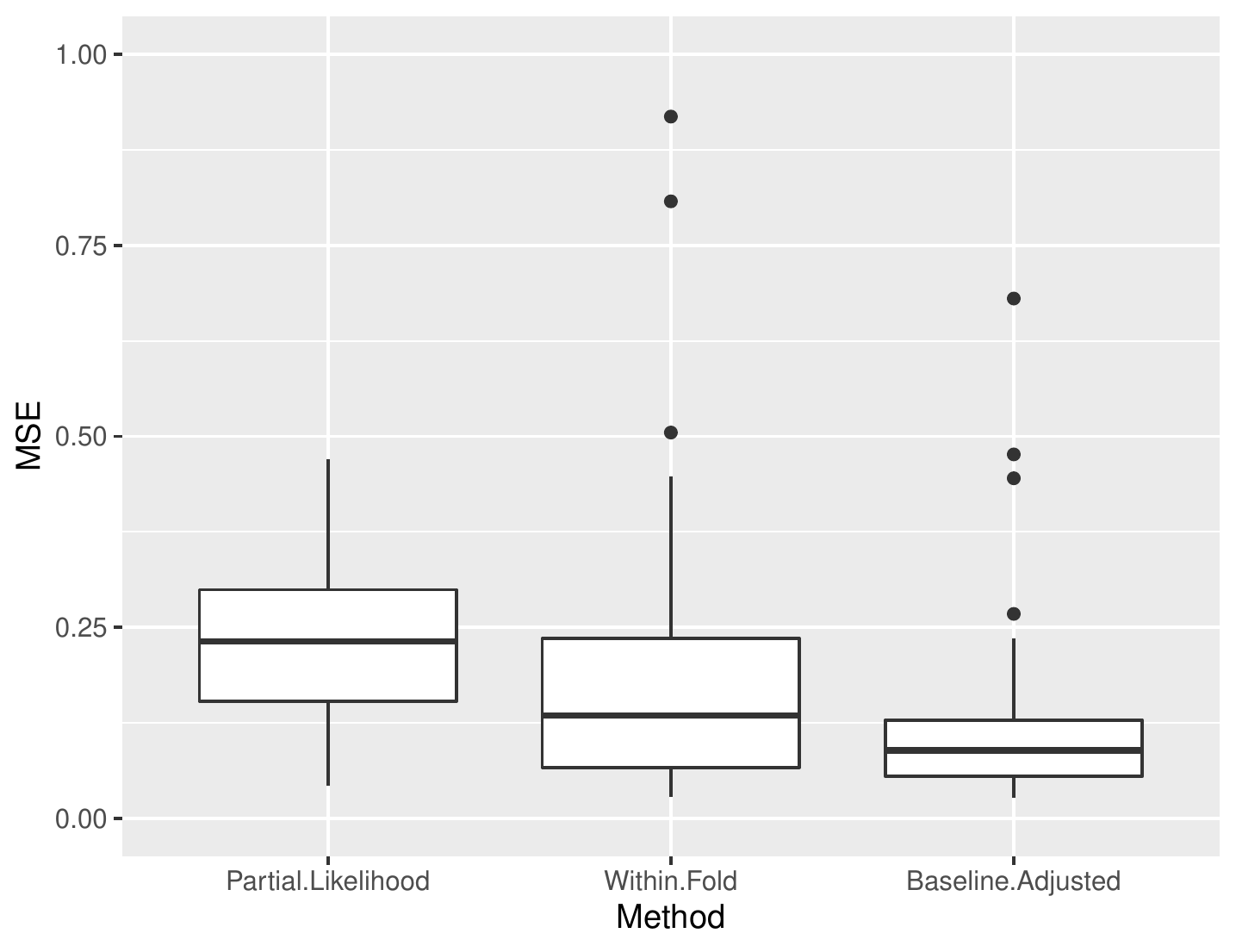} }}%
    \vskip\baselineskip
    \subfloat[Mis-specified Model]{{\includegraphics[width=0.45\linewidth]{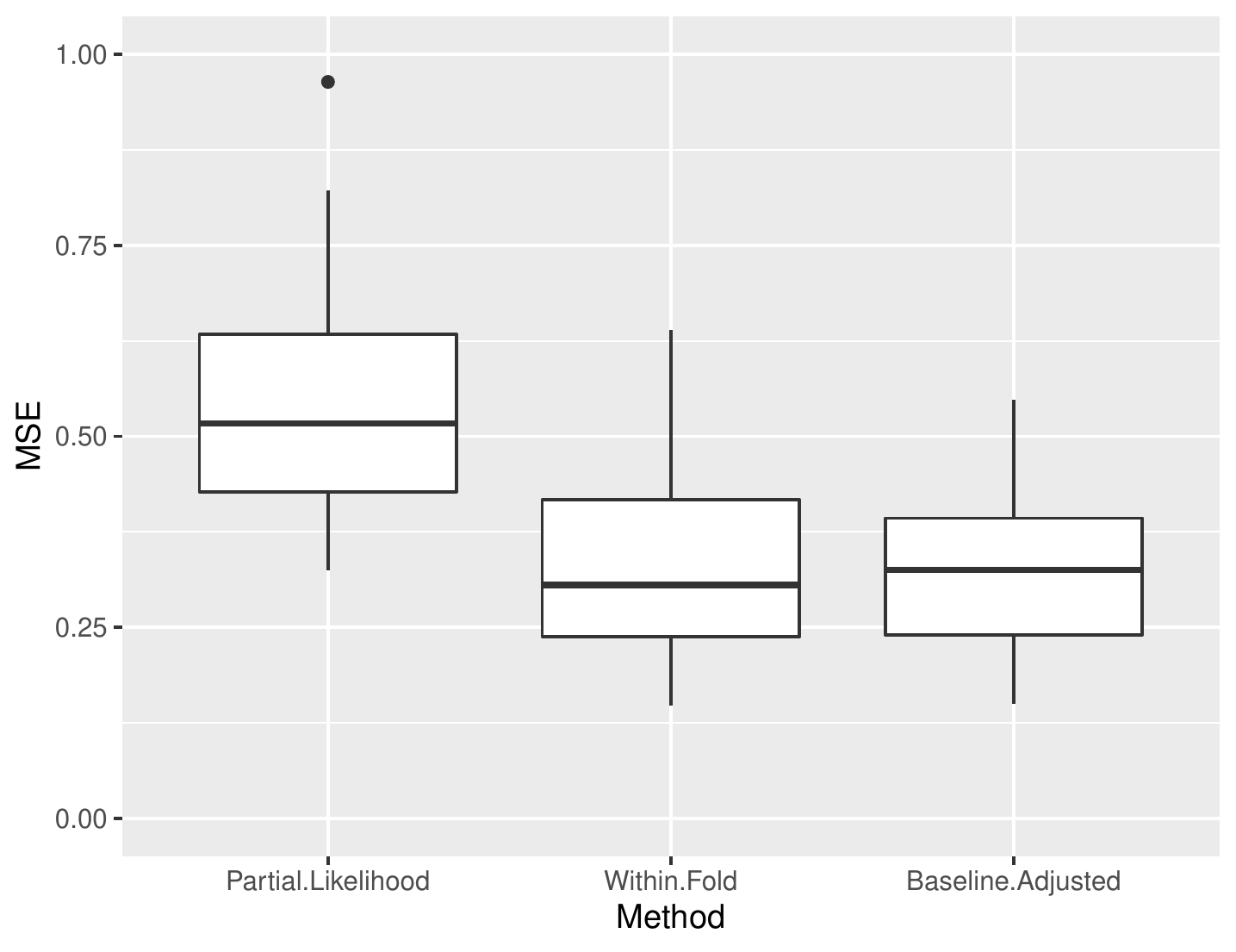} }}%
    \qquad
    \subfloat[AR(1) Predictors, Mis-specified Model]{{\includegraphics[width=0.45\linewidth]{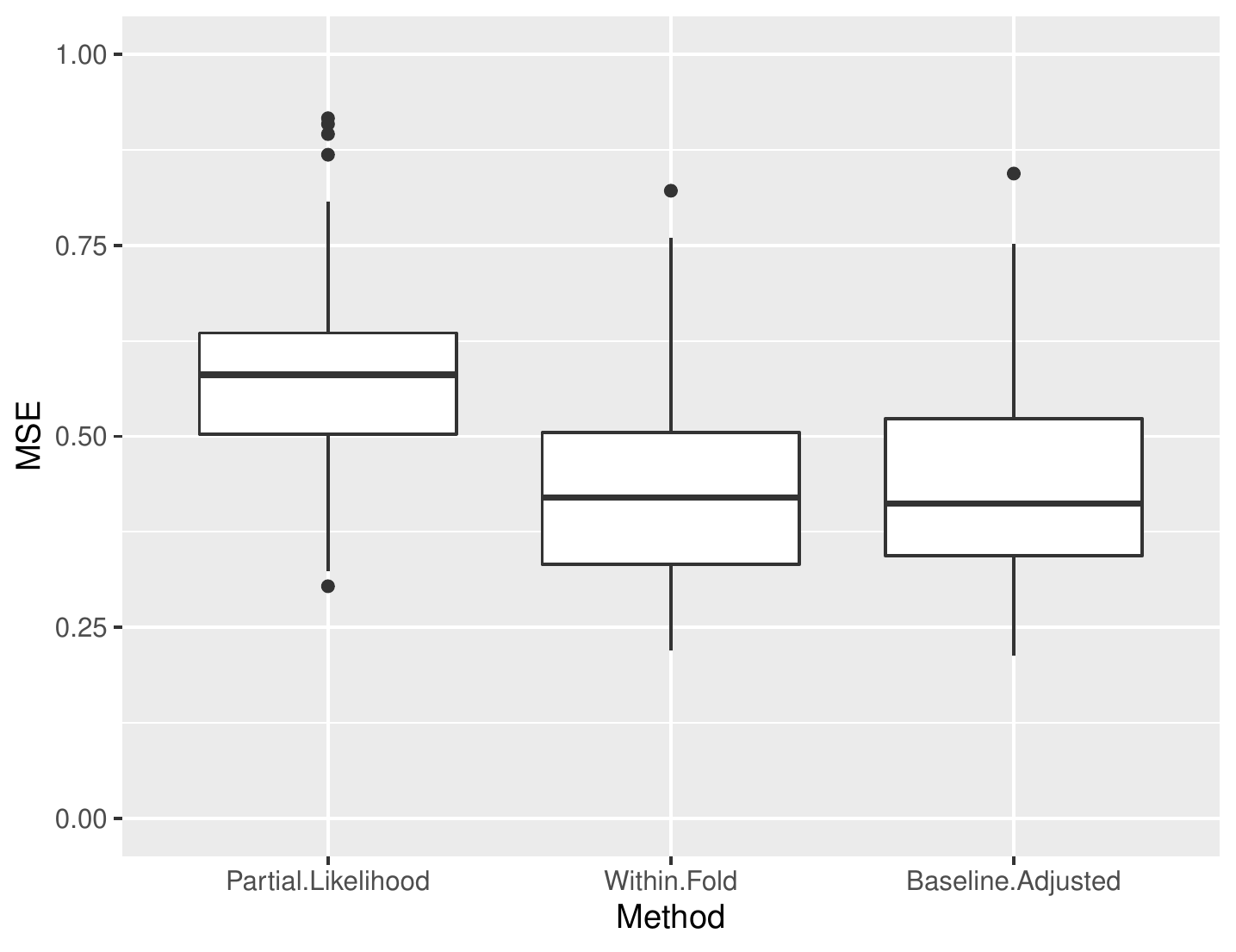} }}%
    
    \caption[]{\em Comparison between three metrics in cross validated Cox model. Each plot describes the mean square error of the fitted parameters using cross validation based on partial likelihood, within fold C-index, and baseline adjusted C-index. Plot(a): Uncorrelated predictors, correctly specified model. Plot(b): AR(1) predictors, correctly specified model. Plot(c): Uncorrelated predictors, mis-specified model. Plot(d): AR(1) predictors, mis-specified model.}%
    \label{FigCV}
\end{figure}
From this plot we see that the baseline adjusted C-index outperforms partial likelihood in all four scenarios, while having similar performance as the within fold C-index in three of the cases. When the predictors follows AR(1) process, the baseline adjusted C-index achieves lower MSE than both partial likelihood and within fold C-index.

\section{Real Data Applications}
We apply our method to the publicly available data set \cite{pmid15374961}, where we use the survival time of the brain cancer patients in the study as the response and their transcriptional profiling as the predictors . Specifically, we first divide the data into training and test subsets. On the training set we do cross-validation where the metric used are partial-likelihood, within-fold C-index, and the baseline adjusted C-index. We then report the C-index of the selected models on the test set. Here is the result:

\begin{figure}[H]
    \centering
    \includegraphics[width=0.8\linewidth]{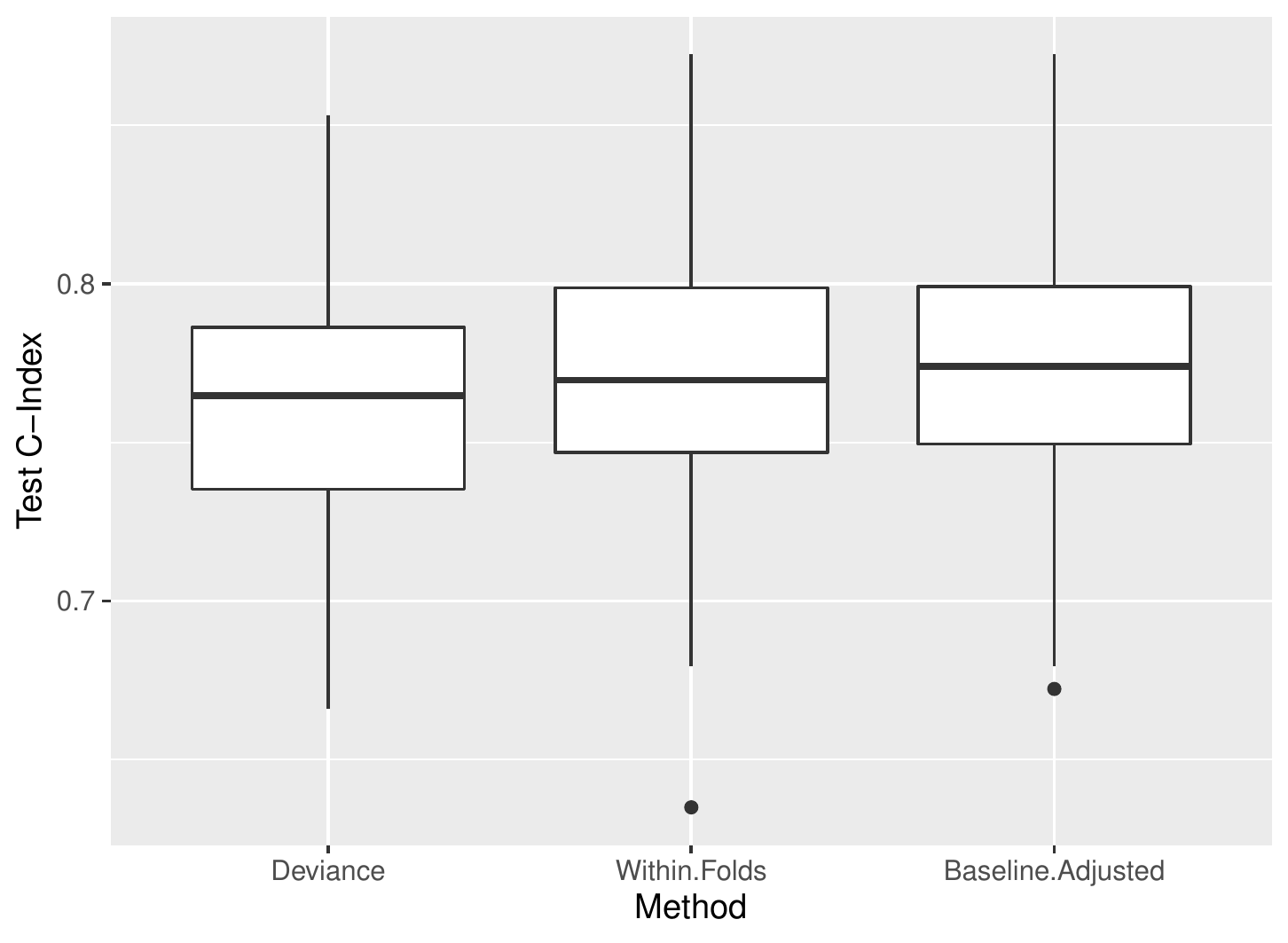}
    \caption{\em{In this experiment we run cross-validation on the brain cancer data. The evaluation metric for CV are deviance, within folds C-index, and the baseline adjusted C-index. For each randomization of train-test-validation split we compute the test C-index using the models selected by these three methods. The box plot is based on $100$ such randomizations.}}
\end{figure}
The above figure shows that on average the baseline adjusted C-index is able to select models that have slightly better test-set C-index than both deviance and the C-index averaged over folds.





\section{Conclusion}
We develop the baseline adjusted C-index to evaluate stratified and cross-validated Cox model. This method makes use of a simple non-parametric estimate of the baseline function according to the parameter estimates and obtains a predicted survival time. Such predicted survival time in turn allows us to compare observations from different strata or cross-validation folds. We demonstrate that the baseline adjusted C-index is a more stable metric of model fit under the stratified Cox model. Under the cross-validation setting, baseline adjusted C-index achieves lower mean-squared error in our simulation, and higher test C-index in the brain cancer dataset.

\bibliographystyle{agsm}
\bibliography{ref}

\end{document}